\documentclass[twocolumn,aip,nofootinbib,jcp,reprint,amsmath,amssymb]{revtex4-1}

\usepackage{natbib,hyperref}
\usepackage{amsmath}
\usepackage{graphicx}
\usepackage{hyperref} 
\usepackage{dcolumn}
\usepackage{bm}
\usepackage{epsfig,color,xspace,multirow,xr,bbold}
\usepackage[all]{xy}
\usepackage{setspace}
\usepackage{url}
\usepackage{soul}
\usepackage[colorinlistoftodos]{todonotes}
\usepackage{longtable}
\usepackage{tabularx}
\usepackage{adjustbox}
\usepackage{threeparttable} 
\usepackage{natbib,hyperref}
\usepackage{float}
\usepackage{placeins}
\usepackage[utf8]{inputenc}
\usepackage[T1]{fontenc}
\usepackage[normalem]{ulem}
\usepackage{xcolor}
\newcommand{\RRef}[1]{Ref.~\onlinecite{#1}}

\newcommand\notsotiny{\@setfontsize\notsotiny\@vipt\@viipt}
\usepackage{soul} % for highlighting
\newcommand{\orcid}[1]{\href{https://orcid.org/#1}{\includegraphics[width=8pt]{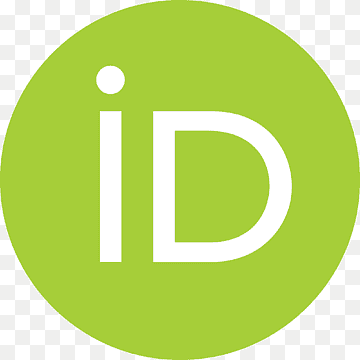}}}
\begin{document}
%\preprint{AIP/123-QED}

\title{
Unlocking Inverted Singlet-Triplet Gap in Alternant Hydrocarbons with 
Heteroatoms 
}
\date{\today}

\author{
Atreyee Majumdar\orcid{0009-0007-4634-5249}, 
Surajit Das\orcid{0009-0001-9339-9430}, 
Raghunathan Ramakrishnan\orcid{0000-0002-7288-9238}
}

\email{ramakrishnan@tifrh.res.in}
\affiliation{$^1$Tata Institute of Fundamental Research, Hyderabad 500046, India}

% \keywords{
% OLEDs,
% Delayed-Fluorescence,
% Singlet-triplet energy gap,
% Heteroaromatics
% }

\begin{abstract}
\noindent
Fifth-generation organic light-emitting diodes exhibit delayed fluorescence even at low temperatures, enabled by exothermic reverse intersystem crossing from a negative singlet–triplet gap (STG), where the first excited singlet lies anomalously below the triplet. 
This phenomenon---termed delayed fluorescence from inverted singlet and 
triplet states (DFIST)---has been experimentally confirmed only in two triangular
molecules with a 12-annulene periphery and a central nitrogen atom.
Here, we report a high-throughput virtual screening of 30,797 BN-substituted polycyclic aromatic hydrocarbons derived from 77 parent scaffolds (2–6 rings). 
Using a multi-level workflow combining structural stability criteria with accurate L-CC2 
excited-state calculations, we identify 72 heteroaromatic candidates with STGs$<0$. 
Notably, this includes BN-helicenes, where inversion arises from through-space charge-transfer states. 
Several systems exhibit non-zero oscillator strengths, supporting their potential as fluorescent emitters.
Our findings reveal new design motifs for DFIST beyond known frameworks, expanding the chemical space for next-generation emitters based on heteroatom-embedded aromatic systems.
\end{abstract}

\maketitle

\section{Introduction\label{sec:intro}}

Thermally activated delayed fluorescence (TADF) enables 100\% exciton harvesting in organic light-emitting diodes (OLEDs), circumventing the dependence on heavy elements to promote phosphorescence\cite{uoyama2012highly}. 
TADF operates via reverse intersystem crossing (RISC), where the population of the `dark' first excited triplet state (T$_1$) is transferred to the  `emissive' higher-energy singlet state (S$_1$), provided the S$_1$-T$_1$ gap (STG) is typically $<0.1$~eV\cite{chen2018thermally}. 
The RISC rate constant, $k_{\rm RISC}$, decreases with temperature ($T$), 
following an Arrhenius-type relation---in the logarithmic form, it is expressed as  
$\ln k_{\rm RISC} = \ln A - {\rm STG}/k_{\rm B}T$,  
where $k_{\rm B}$ is the Boltzmann constant. The intercept, $\ln A$, represents the effective pre-exponential factor within the measured temperature range, while the slope corresponds to the STG divided by $k_{\rm B}$\cite{uoyama2012highly}.
The implication of organic molecules with a negative STG has been 
speculated as $k_{\rm RISC}$ is expected to increase with a decrease in $T$.

Theoretical studies have predicted an inverted STG---{\it i.e.}, a reversal of the typical S$_1$–T$_1$ energy ordering in violation of Hund’s rule---in cyclazine (azaphenalene, 1AP) and heptazine (hepta-azaphenalene, 7AP), both featuring a central nitrogen atom embedded in an anti-aromatic triangular framework derived from a 12-annulene periphery\cite{de2019inverted,ehrmaier2019singlet}.
Aizawa {\it et al.}\cite{aizawa2022delayed} computationally screened approximately 35,000 molecules and identified a 7AP derivative as a promising candidate. 
Experimental measurements revealed a positive slope in the $\ln k_{\rm RISC}$ vs. $1/T$ plot, indicating a negative STG and resulting in delayed fluorescence from inverted singlet and triplet states (DFIST). Molecules exhibiting such behavior are often referred to as INVEST emitters---inverted singlet–triplet energy gap light-emitters---and represent a new frontier in the design of purely organic TADF materials \cite{pollice2021organic}.
Direct spectroscopic evidence of a negative STG ($-0.047 \pm 0.007$ eV) was later obtained for pentaazaphenalene (5AP) using anion photoelectron spectroscopy and fluorescence measurements\cite{wilson2024spectroscopic}, consistent with transient photoluminescence data for dialkylamine-substituted 5AP\cite{kusakabe2024inverted}.
To date, 5AP and 7AP remain the only molecular prototypes with experimentally confirmed negative STGs.

Computational studies have suggested that azaphenalenes 
with other substitution patterns, as well as the boron (B) analog of 1AP,  boraphenalene (1BP), also exhibit negative STGs\cite{won2023inverted,li2022organic,tuvckova2022origin,ricci2021singlet,sancho2022violation,loos2023heptazine,loos2025correction}.
A comprehensive search in the structurally diverse small molecules chemical space, bigQM7$\omega$, \cite{kayastha2022resolution}  
comprising about 13,000 molecules revealed no exceptions to Hund's rule, indicating that achieving the electronic structure criteria for ${\rm STG}<0$ requires non-trivial
molecular frameworks that are inaccessible to molecules with fewer atoms\cite{majumdar2024resilience}. 
Non-alternant hydrocarbons and their substituted analogs have been shown to have the potential to exhibit negative STGs\cite{garner2023double,terence2023symmetry,garner2024enhanced,blaskovits2024excited,nigam2024artificial}. Notably, substituted analogs of the non-fused bicyclic hydrocarbon have demonstrated negative STGs, attributed to through-bond charge-transfer (CT) states\cite{blaskovits2024singlet}. 
The dynamical stability---{\it i.e.}, whether a molecule corresponds to a minimum on the potential energy surface---of novel structural prototypes warrants careful selection of computational protocols, as antiaromatic frameworks combined with topologically charge destabilizing substitutions
are particularly susceptible to Jahn--Teller-type distortions, affecting the predicted STG\cite{majumdar2024influence}.
\begin{figure*}[hptb]
    \centering
    \includegraphics[width=\linewidth]{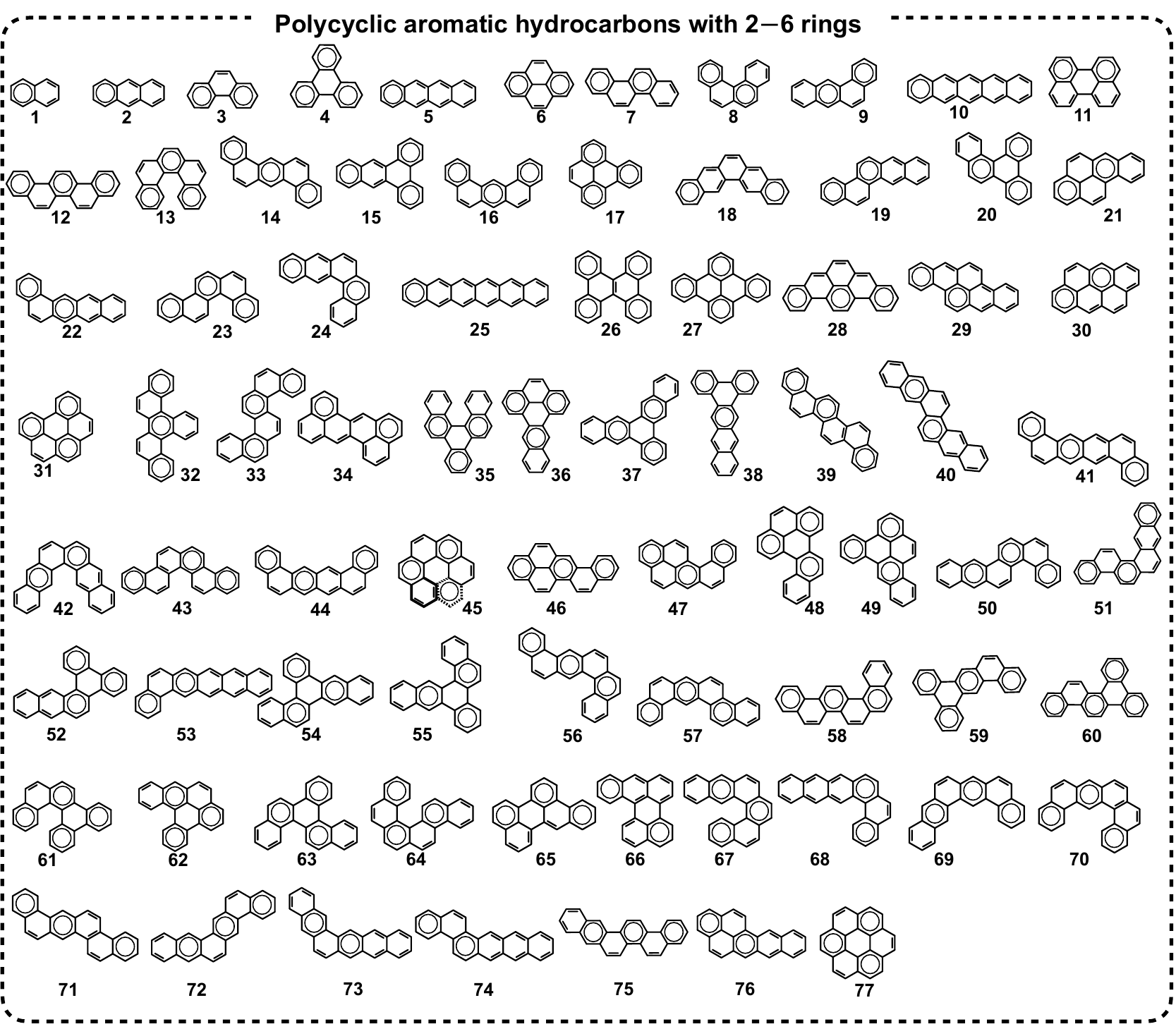}
    \caption{
    \small
Set of 77 smallest polycyclic aromatic hydrocarbons (PAHs) comprising up to six benzene rings, adapted from \RRef{chakraborty2019chemical}. These serve as parent scaffolds for the BNPAH chemical space explored in this work. Molecules include linear, angular, and helical topologies.
Names and SMILES are collected in Table~S1 of the Supplementary Information (SI). 
\label{fig:PAH}
}
\end{figure*}

Heteroatom-substituted polycyclic aromatic hydrocarbons (PAHs), particularly those incorporating B and N (BNPAH), constitute a vast chemical space\cite{chakraborty2019chemical} with significant potential for developing versatile organic semiconductors \cite{matsui2018one,borissov2021recent,chen2023boron}. 
An emerging strategy is to enhance the efficiency of TADF-based OLEDs with multiple-resonance fluorophores, as in the case of DABNA, a BNPAH molecule with a triphenyl boron core and two N atoms\cite{hatakeyama2016ultrapure}.
BNPAH with separated B and N centers present the possibility of
through-bond CT states which can result in a negative STG as noted in calicene\cite{blaskovits2024singlet}.
For instance, derivatives of tetracene\cite{pimentel2025thermally} and Clar's goblet diradical\cite{derradji2025functionalization} exhibit small positive STGs 
highlighting their potential for TADF applications. 
However, so far DFIST candidates with STG$<0$, have not been identified in the BNPAH chemical space, as  the key challenge lies in not only selecting an appropriate PAH scaffold but also the suitable substitution pattern.
To identify such rare candidates, high-throughput, \textit{ab initio} screening is key---enabling comprehensive exploration of the expansive BNPAH chemical space to uncover rare scaffold motifs and heteroatom substitution patterns that meet the stringent electronic structure requirements for achieving inverted STGs.

In this study, we comprehensively explore DFIST candidates with ${\rm STG}<0$ within a chemical space of 30,797 BNPAH molecules\cite{chakraborty2019chemical} of stoichiometry C$_x$B$_1$N$_1$. These molecules were combinatorially derived from 77 benzenoid Kekul\'ean PAHs containing 2–6 rings\cite{brinkmann2007fusenes}, including coronene (superbenzene). The structures of 
these PAHs are shown in Figure~\ref{fig:PAH}.
By applying stringent filters accounting to filter systems prone to de-excitations and structural distortions, we identified 46 stable systems with negative STGs. Notably, the most prevalent candidates were non-planar helicene-type structures and azaphenalene and boraphenalene-based systems. A detailed analysis of these structures revealed key features contributing to their negative singlet-triplet gaps, offering new directions for designing organic emitters with efficient DFIST properties.

\section{Computational details\label{sec:methods}} 
\begin{figure*}[hptb]
    \centering
    \includegraphics[width=\linewidth]{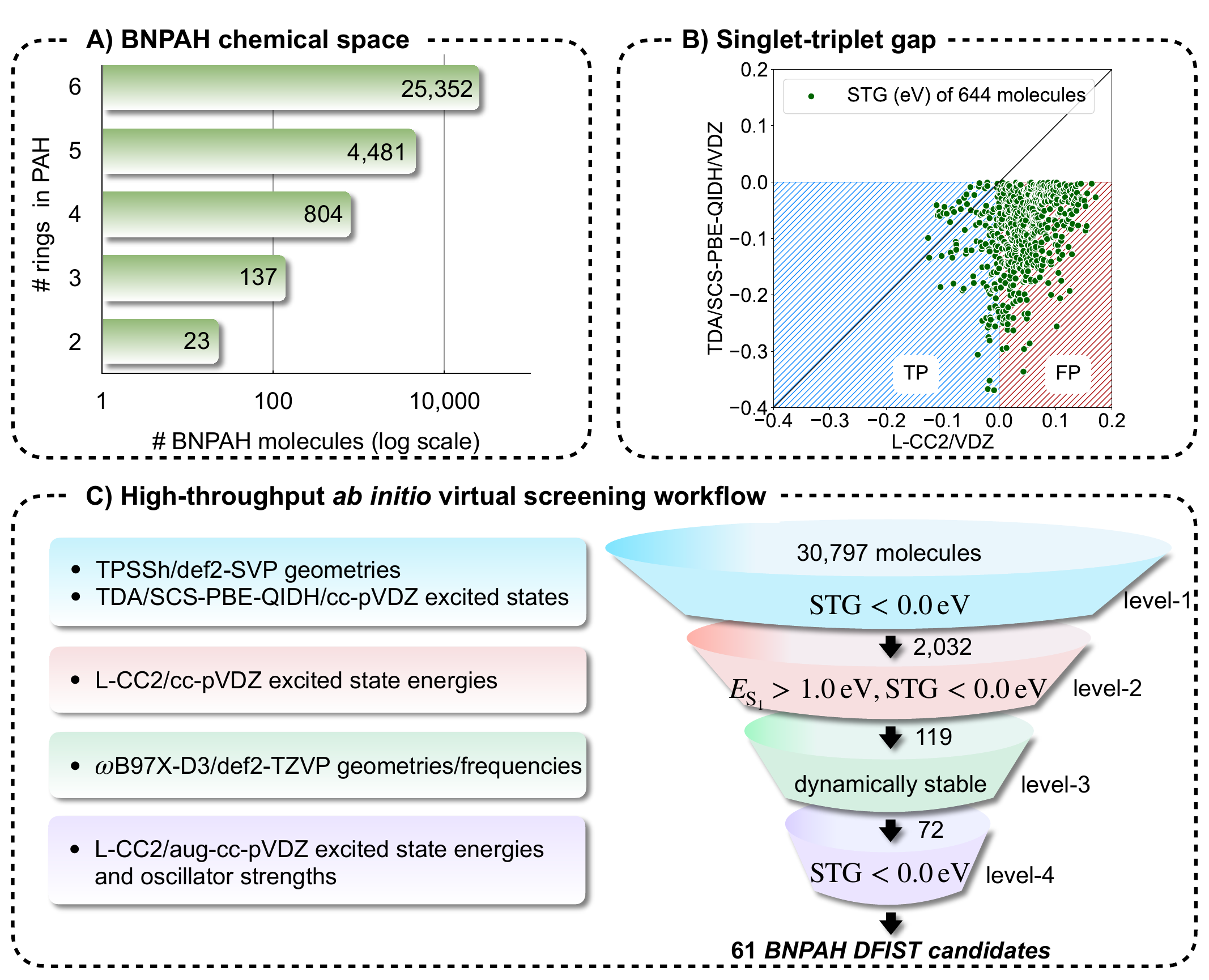}
    \caption{
Data distribution and high-throughput virtual screening workflow:
(A) Distribution of 30,797 BNPAH molecules categorized by PAH size in terms of the number of rings.
(B) Correlation between STGs of 644 BNPAH molecules predicted by TDA/SCS-PBE-QIDH and L-CC2 methods. Out of the 2,032 molecules with negative STGs at the DFT level, 644 molecules with S$_1$ energies $>1$ eV at the L-CC2 level are retained for further analysis. TP (true positives) and FP (false positives) indicate the classification accuracy of TDA/SCS-PBE-QIDH in predicting negative STGs, with L-CC2 serving as the reference. 
(C) High-throughput workflow for identifying DFIST candidates with reliable minimum-energy geometries and accurate excited-state characteristics.
\label{fig:workflow}
}
\end{figure*}

The geometries of 30,797 BNPAH molecules 
derived from 77 PAHs
were obtained using density functional theory (DFT) with the TPSSh exchange-correlation (XC) functional and the def2-SVP basis set, as reported in \RRef{chakraborty2019chemical}. 
We note that for helical PAHs that are chiral, only one enantiomeric
form is considered in
\RRef{chakraborty2019chemical}. 
Screening for DFIST candidates within this dataset is carried out using the computational workflow illustrated in Figure~\ref{fig:workflow}, which consists of four levels. At level-1, the initial screening of DFIST candidates, is performed using linear-response time-dependent DFT (LR-TD-DFT) calculations within the Tamm–-Dancoff approximation (TDA) using the SCS-PBE-QIDH double hybrid DFT method\cite{casanova2021time} combined with the correlation-consistent polarized valence double-zeta (cc-pVDZ) basis set. Molecules with STG $<0$ at this stage progress to the next level of the workflow for further evaluation.

The large size of the BNPAH molecules considered in this study limits the feasibility of using canonical ADC(2) or CC2 methods at level-2 of the workflow. In contrast, the local L-CC2 method provides a significant computational speed-up without compromising accuracy. Therefore, level-2 employs the Laplace-transformed, density-fitted, local variant of the second-order approximate coupled-cluster singles and doubles method (L-CC2)\cite{freundorfer2010local}, using the cc-pVDZ basis set along with the corresponding JKFIT and MP2FIT auxiliary basis sets. Molecules with STG $<0$ as predicted by L-CC2/cc-pVDZ proceed to level-3 of the workflow, where  
the geometries of the candidate molecules are refined using the DFT method $\omega$B97X-D3 with the def2-TZVP basis set. Vibrational analysis at the same level is performed to assess their dynamic stability, determining whether they correspond to minimum-energy structures on the potential energy surface (PES). 
At level-4, excited-state calculations are performed using L-CC2 with the larger aug-cc-pVDZ basis set, utilizing the refined geometries of the dynamically stable molecules obtained from $\omega$B97X-D3/def2-TZVP. Molecules that pass this final stage are identified as DFIST candidates with a higher degree of confidence.

Multiwfn\cite{lu2012multiwfn} was used for calculating $\Lambda$-indices for which excited states calculations were performed with 
TDA/SCS-PBE-QIDH/cc-pVDZ. Molecular orbitals generated in these TDA calculations were plotted.
In both TDA and L-CC2 calculations, we evaluated the lowest six singlet and triplet excited states. All L-CC2 calculations were carried out using Molpro (version 2015.1)\cite{werner2015molpro} and
all DFT calculations were performed using Orca (version 6.0.0) \cite{neese2012orca, neese2018software}. 

To assess the aromatic character of the rings in selected BNPAH molecules, we employed the harmonic oscillator model of aromaticity (HOMA) \cite{kruszewski1972definition}. For a given $(4n)$ or $(4n+2)\pi$ H\"uckel-type ring, the HOMA index is defined as
\begin{equation}
    {\rm HOMA} = 1 - \frac{1}{N} \sum_i^{N} \alpha^{\rm AB}_i (R_{i, {\rm opt}}^{\rm AB}-R_{i, {\rm calc.}}^{\rm AB} )^2,
\end{equation}
where $N$ is the number of bonds in the ring, $R_{i, {\rm calc.}}^{\rm AB}$ is the DFT-predicted bond length for each A–B type bond, $R_{i, {\rm opt}}^{\rm AB}$ is the optimal bond length, and $\alpha^{\rm AB}_i$ is an empirical constant chosen such that a fully delocalized aromatic ring yields HOMA = 1. A HOMA value near 1 indicates strong aromaticity, while values below 0.5 suggest significant deviation from aromatic behavior and increased bond-length alternation. 
Non-aromatic Kekul{\'e}-like structures typically yield HOMA $\approx0$, and antiaromatic systems can have negative HOMA values. 
For CC/CN/CB/BN bonds, we used $R_{\rm opt}^{\rm AB}=1.388/1.334/1.4235/1.402$ \AA{} 
and $\alpha^{\rm AB}=257.7/93.52/104.507/72.03$ \AA$^{-2}{}$\cite{krygowski1993crystallographic,zborowski2012calculation,madura1998structural,pimentel2025thermally}.

\section{Results and Discussions\label{sec:results}} 
The results of this study are presented in three parts.
(i) The first part discusses the salient aspects---organization of different levels, selection of the methods employed---of the computational workflow used for 
high-throughput screening of DFIST molecules.
(ii) Structure-property correlation between L-CC2/cc-pVDZ-level STGs and the molecular
structural features for a rational understanding of the origin of the property. 
We analyze 644 molecules (with S$_1$ energies $>1$ eV) out of 2,032 passing level-1 
(Figure~\ref{fig:workflow}) of the workflow with both positive and negative STGs. 
(iii) Excited state characteristics of DFIST candidates passing the final level of 
the workflow modeled with L-CC2/aug-cc-pVDZ using refined geometries calculated 
with $\omega$B97X-D3/def2-TZVP.

\subsection{Assessment of data quality\label{label:subsec1}}
We begin with the geometries of 30,797 BNPAH molecules, calculated using the TPSSh-DFT method with the def2-SVP basis set, as reported in \RRef{chakraborty2019chemical}. The computational workflow employed in this study (see Figure~\ref{fig:workflow}) systematically identifies DFIST candidates by progressively increasing theoretical rigor at each stage. The primary objective of the workflow is to ensure that the final set of identified candidates are true positives, meaning they meet the required property criteria. While some true positives may be missed due to the sequential filtering steps, the workflow prioritizes minimizing false positives, ensuring high confidence in the candidates that successfully pass all levels of screening.

An ideal first step in the workflow would involve refining the geometries of BNPAH molecules using a more accurate DFT method, especially for geometry optimization, such as $\omega$B97X-D3, with a larger def2-TZVP basis set. However, the computational cost of geometry refinement and vibrational frequency analysis for a dataset of this scale using a triple-zeta basis set is prohibitively high. Therefore, geometry refinement is deferred to the third stage of the workflow and applied only to candidate molecules that successfully pass the first two screening levels.

% Previous studies have shown that in certain azaphenalenes, the inverted nature of STG is highly sensitive to the choice of geometry optimization methods. To address this, we refine geometries using the more accurate $\omega$B97X-D3 functional and a larger def2-TZVP basis set. 

Level-1 of the computational workflow performs initial screening of  30,797 BNPAH molecules derived from the smallest 77 PAHs (see Figure~\ref{fig:PAH}) using excited state energies calculated at the TDA-SCS-PBE-QIDH/cc-pVDZ level.
TDA/SCS-PBE-QIDH method has shown good agreement with theoretical best estimates for predicting negative STGs in azaphenalenes\cite{loos2023heptazine}. While various double-hybrid density functional approximations (dh-DFAs) achieve either very low mean absolute deviations (MADs) or low standard deviations of errors (SDEs), SCS-PBE-QIDH consistently maintains low values for both metrics\cite{majumdar2025leveraging}. More importantly, previous studies have shown that TDA/SCS-PBE-QIDH systematically underestimates STG compared to more accurate reference methods \cite{loos2023heptazine,majumdar2024resilience}.
Thus, the probability of false-negative predictions of inverted STGs with this method is very low. Consequently, molecules identified with positive STGs at this level are highly unlikely to exhibit negative STGs when evaluated with higher-level theories. From the initial set of 30,797 BNPAH molecules, we filtered out those exhibiting de-excitations---negative transition energies for S$_1$ or T$_1$---indicating a preference for open-shell singlet or triplet electronic ground state rather than a closed-shell singlet. 
In the remaining 30,319 molecules, we selected 
those with STG $< 0.0$ eV identifying 2,032 candidate molecules for subsequent assessment. 

At level-2 of the workflow, excited-state energies are computed for 2,032 molecules using the L-CC2 method with the cc-pVDZ basis set. L-CC2 has demonstrated high accuracy in modeling excited states of BODIPY derivatives\cite{feldt2021assessment, momeni2016local}. 
For 12 azaphenalenes, L-CC2 with the aug-cc-pVTZ basis set achieves mean absolute deviations (MAD) and standard deviations of errors (SDE) of 0.015 and 0.010 eV (see S3 of SI), respectively\cite{majumdar2025leveraging}, relative to theoretical best estimates from \RRef{loos2023heptazine} (see S2 of SI). For the same set, the canonical CC2 method with the aug-cc-pVTZ basis set, as reported in \RRef{loos2023heptazine}, yields MAD and SDE values of 0.013 and 0.011 eV (see S3 of SI), respectively. These results indicate that the local approximation in L-CC2 does not degrade predictive accuracy.
In this study, we screen 2,032 molecules using L-CC2 with the smaller cc-pVDZ basis set. For 12 azaphenalenes, L-CC2/cc-pVDZ achieves MAD and SDE of 0.039 and 0.020 eV compared to theoretical best estimates (see S4 of SI). \RRef{majumdar2025leveraging} further demonstrated that L-CC2's accuracy with the aug-cc-pVDZ basis set closely matches that of the aug-cc-pVTZ basis set, with MAD and SDE values of 0.016 and 0.013 eV, respectively (see S4 of SI). 
%A summary of L-CC2 accuracy is provided in Table~S4 of the SI.

Based on L-CC2/cc-pVDZ results for the 2,032 molecules, we excluded 10 molecules exhibiting de-excitations and 1,388 molecules with $E_{{\rm S}_1} > 1$ eV, as they are prone to pseudo-Jahn–Teller distortions. This resulted in a final set of 644 molecules. The 1 eV transition energy threshold was chosen based on previous findings on azaphenalenes, which showed increasing pseudo-Jahn–Teller distortions with decreasing ${\rm S}_0\rightarrow{\rm S}_1$ transition energy\cite{majumdar2024influence}. However, the likelihood of structural distortions also depends on the strength of vibronic coupling between the corresponding states.
We analyzed the full set of 644 molecules, even though some exhibit STG $>0$, as their STG magnitudes remain small due to prior screening at level-1. This broader structural analysis facilitates the identification of trends, such as the dependence of STG magnitude and sign on the choice of PAH scaffold and other key structural features.

Of the 644 molecules entering level-2, we focus on 119 molecules with STG $< 0$ for further analysis in level-3, where their geometries are refined using the DFT method $\omega$B97X-D3 with the def2-TZVP basis set. 
Vibrational analysis confirmed that 72 molecules are dynamically stable, meaning they correspond to minimum-energy structures on the PES. This set includes six molecules with a soft mode, where the corresponding imaginary frequency magnitude is below 1 terahertz (1 THz = 33.356 cm$^{-1}$).

The choice of $\omega$B97X-D3 for geometry refinement is motivated by its demonstrated reliability in large computed datasets, such as the QM9 chemical space\cite{ramakrishnan2014quantum}, where $\omega$B97-X-D3 has been shown to outperform B3LYP in predicting minimum-energy structures for molecules with unusual covalent bond connectivities, many of which have yet to be experimentally identified\cite{senthil2021troubleshooting}.
Furthermore, for azaphenalenes, $\omega$B97X-D3 accurately predicts vibronic coupling-driven pseudo-Jahn–Teller symmetry-lowering in agreement with the coupled-cluster CCSD(T) method\cite{majumdar2024influence}. In contrast, the CCSD method results in excessive symmetry breaking, while commonly used geometry optimization methods—B3LYP hybrid-DFT and second-order perturbation theory (MP2)—fail to adequately capture vibronic coupling effects, incorrectly predicting symmetric structures as minima that are actually saddle points at the CCSD(T) level\cite{majumdar2024influence}.

Finally, at level-4, excited-state calculations for the final set of 72 stable molecules were performed using L-CC2 with the larger aug-cc-pVDZ basis set, utilizing the refined geometries obtained with $\omega$B97X-D3/def2-TZVP.

 \begin{figure}[htp]
    \centering
    \includegraphics[width=\linewidth]{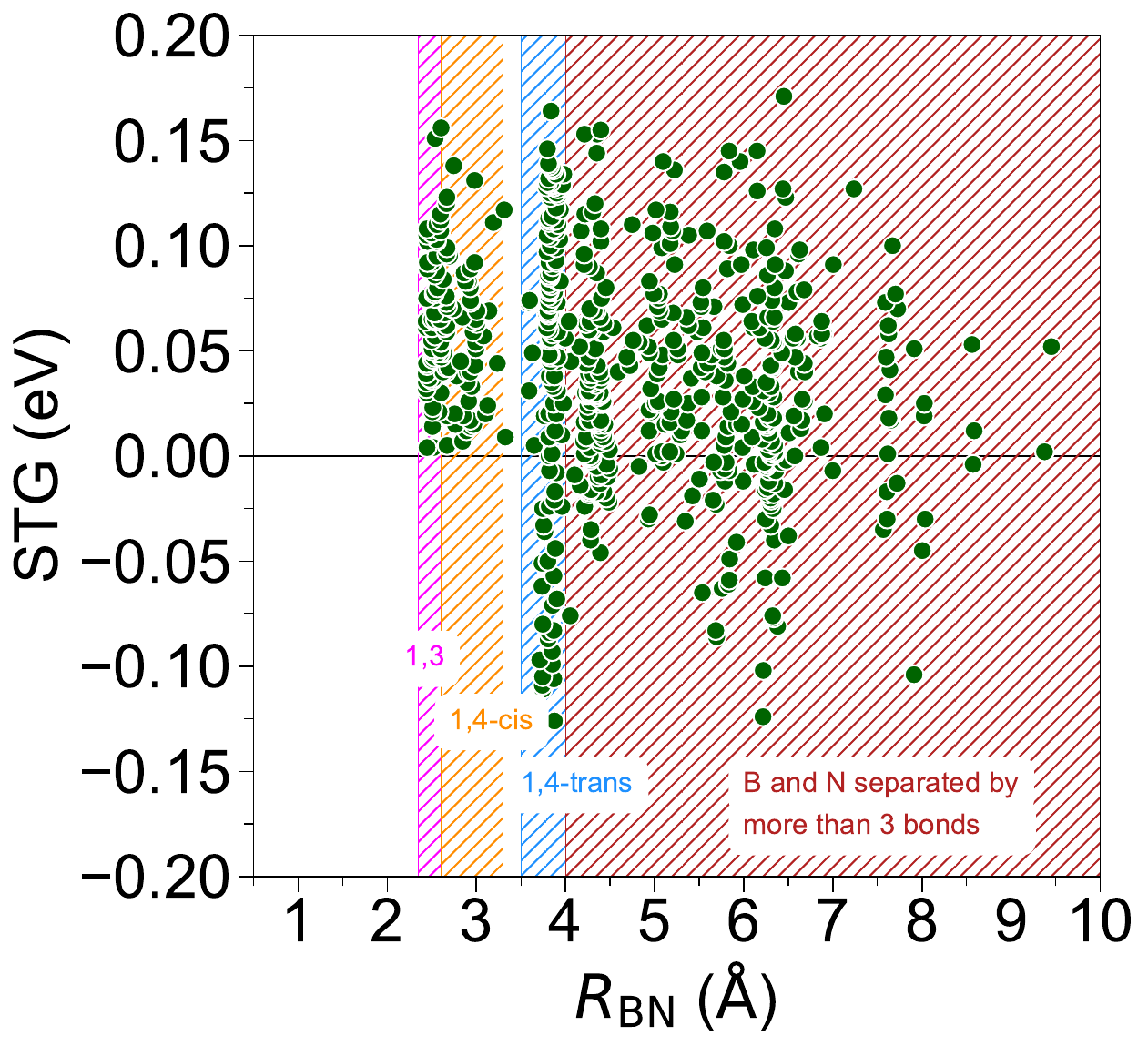}
    \caption{
Dependence of L-CC2 STGs on the distance between B and N 
atoms in 644 BNPAH molecules. Typical distances for 2-bond separations 
(1,3) are shaded in magenta, while 3-bond separations are distinguished 
as 1,4-cis and 1,4-trans (analogous to cis/trans butadiene) and shaded 
in orange and blue, respectively. Distances for beyond 3-bond separations 
are shaded in red.
\label{fig:RBNSTG}
}
\end{figure}

\begin{figure*}[htp]
    \centering
    \includegraphics[width=\linewidth]{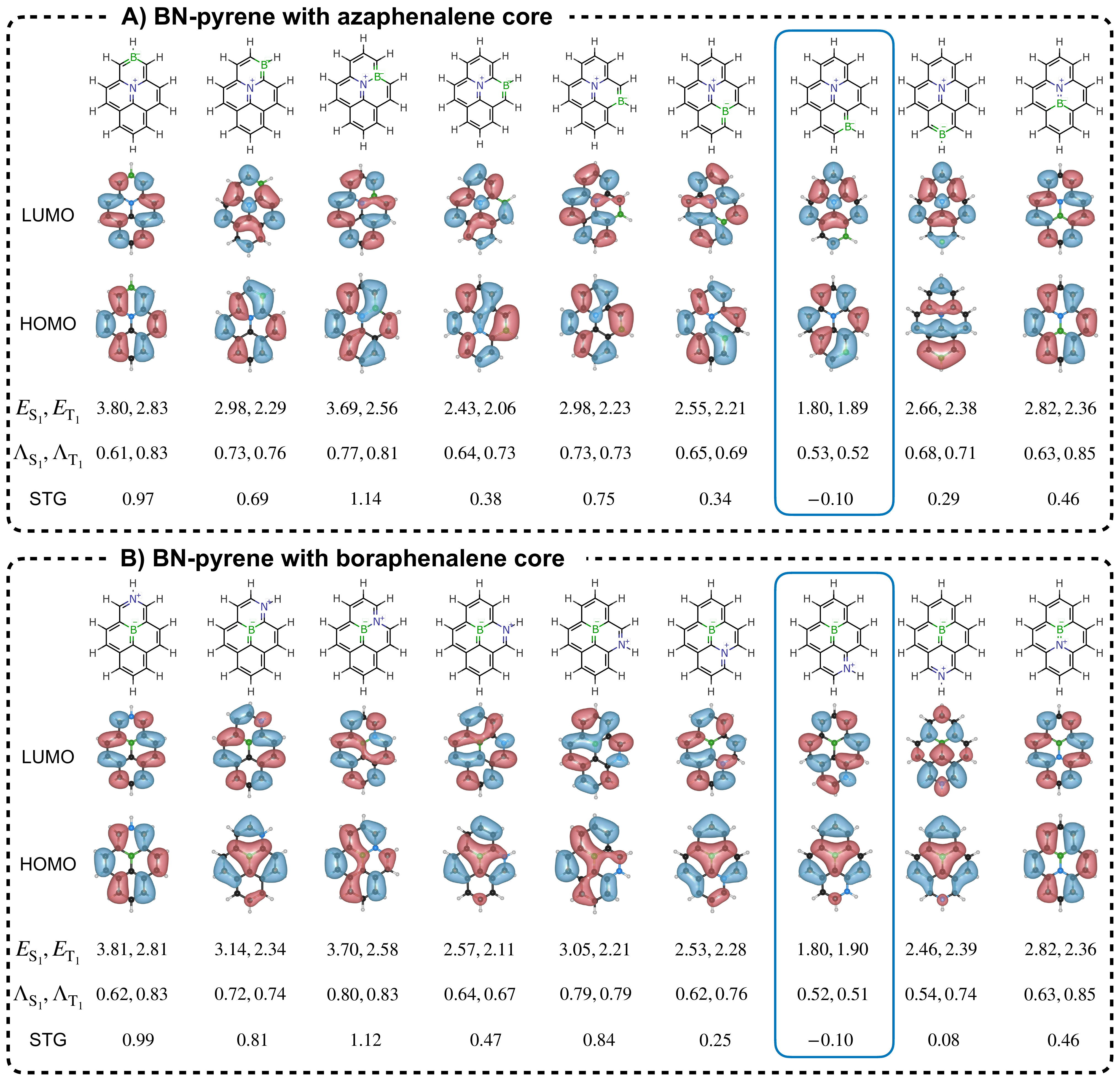}
    \caption{
B,~N-substituted pyrenes with azaphenalene and boraphenalene cores:
Variation of frontier MOs shown along with the L-CC2/cc-pVDZ 
transition energies of S$_1$ and T$_1$ states along with STG (in eV). 
Also given are the $\Lambda$-indices for S$_1$ and T$_1$ transitions 
and the corresponding frontier MOs calculated with 
TDA-SCS-PBE-QIDH/cc-pVDZ. 
\label{fig:BNpyrene}
}
\end{figure*}

\subsection{Dependence of singlet-triplet gap on structural factors\label{label:subsec2}}

The analysis in this subsection is based on the 644 molecules identified at level-2 of the workflow. As noted earlier, these molecules have $E_{{\rm S}_1} > 1$ eV at the L-CC2/cc-pVDZ level 
(using TPSSh/def-SVP minimum energy geometries retrieved from the BNPAH dataset reported in \RRef{chakraborty2019chemical})
and include 525 molecules with positive STGs and 119 molecules with negative STGs.

 \begin{figure*}[htpb]
    \centering
    \includegraphics[width=\linewidth]{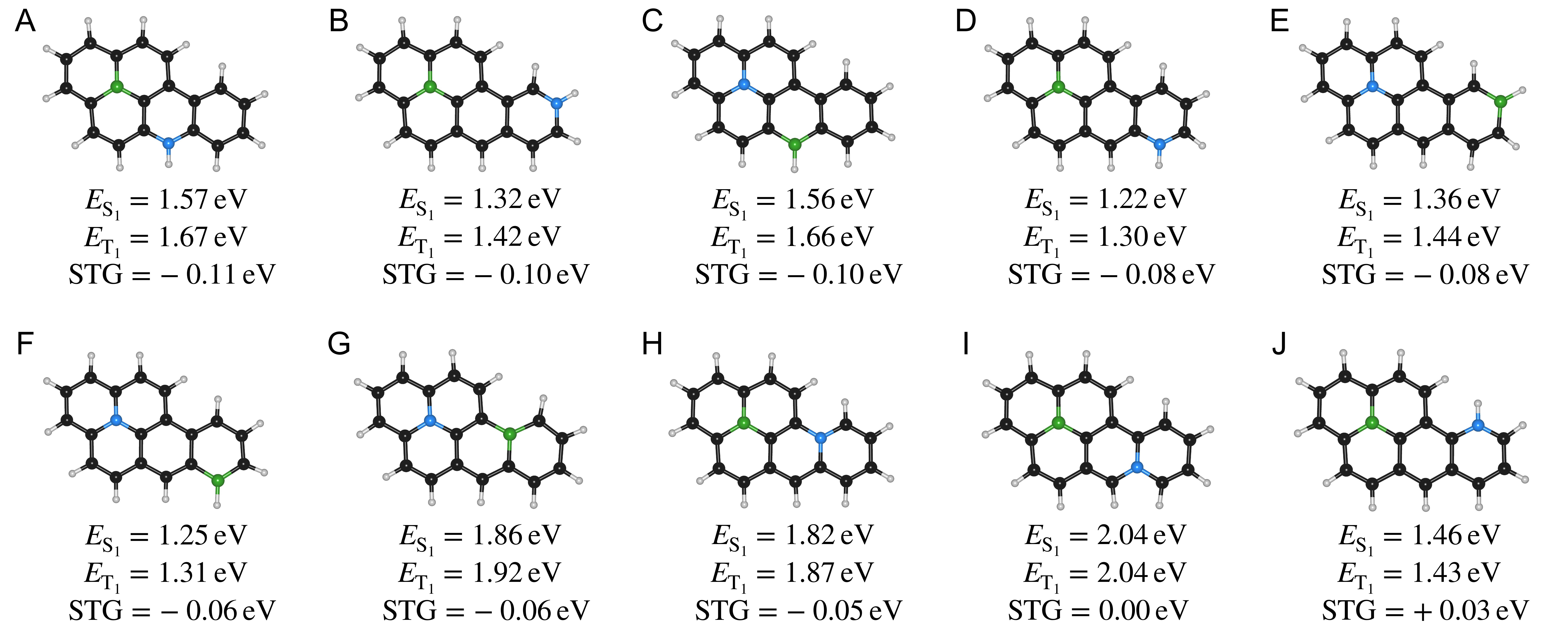}
    \caption{
Structures and L-CC2/cc-pVDZ-level excited-state properties of 10 BN-benzo[a]pyrene molecules derived from PAH \#21 (see Figure~\ref{fig:PAH}). White$|$green$|$black$|$blue atoms denote H$|$B$|$C$|$N. 
\label{fig:PAH21} 
}
\end{figure*}

\paragraph{Variation of STG with B-N separation:}
 Figure~\ref{fig:RBNSTG} illustrates the variation of L-CC2 predicted STG with B–N distance ($R_{\rm BN}$) 
 for 644 BNPAH molecules. 
Notably, molecules with directly bonded B and N atoms are absent from this plot, as they exhibited STG $> 0.1$ eV at the SCS-PBE-QIDH level (did not pass level-1 of the workflow Figure ~\ref{fig:workflow}) 
and were therefore excluded from subsequent modeling with L-CC2. 
Additionally, molecules with B and N separated by two bonds, similar to 
meta-positions in benzene (indicated as 1,3 in Figure~\ref{fig:RBNSTG}), consistently show STG $> 0$ eV, indicating that such configurations cannot support negative STGs. 
This is likely due to the shorter B–N distances increasing the overlap of frontier molecular orbital densities involved in the excitations, which in turn enhances exchange interactions and stabilizes ${\rm T}_1$ relative to ${\rm S}_1$. 
Similarly, negative STGs are not observed in systems where B and N are separated by two bonds, as seen in para-positions in benzene or the 1,4 positions of cis-butadiene (denoted as 1,4-cis in Figure~\ref{fig:RBNSTG}). 
However, for the 1,4-trans configuration, several molecules exhibit negative STGs, suggesting that an increase in through-space B–N distance plays a crucial role in minimizing exchange interactions, thereby enabling negative STGs.
No particular selectivity for negative STG is noted in Figure~\ref{fig:RBNSTG} for systems with $R_{\rm BN}>4$~\AA{} based on the BNPAHs explored in this study.

\begin{figure*}[htpb]
    \centering
    \includegraphics[width=\linewidth]{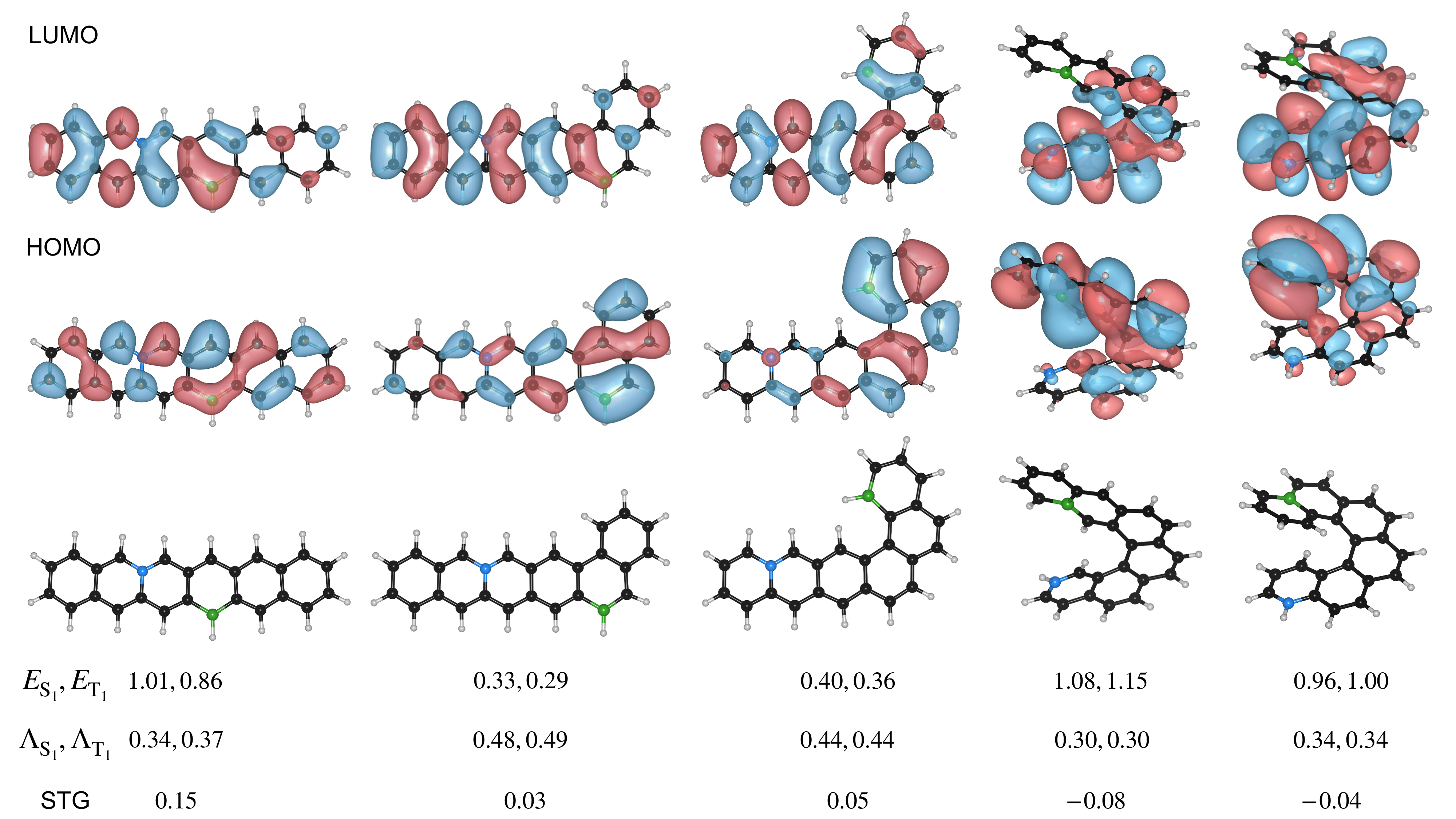}
    \caption{
Variation of STG in 6-ring BNPAHs with increasing helicity. 
For each PAH scaffold, the molecule with the smallest STG at the L-CC2/cc-pVDZ level is shown. 
L-CC2 transition energies and STG values are provided in eV, along with $\Lambda$-indices and frontier MOs calculated using TDA-SCS-PBE-QIDH/cc-pVDZ. White$|$green$|$black$|$blue atoms denote H$|$B$|$C$|$N. 
\label{fig:BNhelicenes} 
}
\end{figure*}
%Analyse the series 25-53-68-67-45
% 7849,18048,27876,27085,13516

\paragraph{BN-pyrene molecules with 1AP or 1BP core:}
To gain a more refined understanding of the qualitative trends, we selected all BN-pyrene molecules containing either a 1AP or a 1BP core, from a total of 63 possible BN-pyrenes\cite{chakraborty2019chemical}, 
as illustrated in Figure~\ref{fig:BNpyrene}. 
Each case consists of nine molecules, among which seven feature either directly bonded B \& N pairs, 2-bond-separated 1,3 configurations, or 3-bond-separated 1,4-cis connectivities---none of which favor negative STGs as seen from the trends in Figure~\ref{fig:RBNSTG}. 
The remaining two systems in each set have larger B–N distances ($R_{\rm BN}$): one exhibits a 3-bond-separated 1,4-trans configuration, while the other has a 4-bond separation. 
In both classes of BN-pyrenes shown in Figure~\ref{fig:BNpyrene}, the left and right most entries wih a two-fold symmetry and short B-N distances, the frontier MOs resemble that of the aromatic species 14-annulene with no contribution from the heteroatoms. 
The frontier molecular orbitals (MOs) of the 1,4-trans systems closely resemble those of 1AP and 1BP discussed in \RRef{loos2023heptazine}. For the 4-bond-separated system with an 1AP core, the LUMO is similar to that of 1AP, but the HOMO exhibits different nodal structures. In contrast, for the 4-bond-separated system with a 1BP core, the HOMO resembles that of 1BP, whereas the LUMO does not.
Additionally, while both systems show reduced STGs, the 1,4-trans systems (highlighted in boxes) exhibit slightly lower $\Lambda_{{\rm T}_1}$ compared to $\Lambda_{{\rm S}_1}$, suggesting diminished exchange energy in the triplet state, which 
favors a lower-energy singlet state.
Overall, these results suggest that in B,~N-substituted pyrenes, negative STG does not solely stem from the 1AP or 1BP core but also requires B and N to be connected in a 1,4-trans butadiene-like fashion.

\paragraph{BN-benzo[a]pyrene molecules:}

From the 644-set, we selected 10 BNPAH molecules derived from PAH \#21, benzo[a]pyrene (see Figure~\ref{fig:PAH}). Their structures, along with L-CC2/cc-pVDZ excited-state energies, are shown in Figure~\ref{fig:PAH21}. All 10 BN-benzo[a]pyrene molecules feature either a 1AP or 1BP core. Note that the full set of BN-benzo[a]pyrene derivatives comprises 380 unique molecules\cite{chakraborty2019chemical}, many of which were filtered out at level-1 of the workflow based on STG thresholds and stability criteria.

Among them, eight molecules exhibit negative STG (excluding I and J) and are related by permutational symmetry. Of these, four (A, B, D, and H) contain a 1BP core, while the other four (C, E, F, and G) are obtained by swapping the positions of B and N. Furthermore, within these eight systems with STG$<0$, four (A, C, G, and H) have B and N connected via a 1,4-trans configuration. In A and C, the peripheral B and N atoms are bonded to hydrogen atoms, whereas in G and H, they occupy bridgehead positions. Overall, while 1,4-trans connectivity promotes negative STG, the presence of a heteroatom at a bridgehead position increases the individual transition energies ($E_{{\rm S}1}$ and $E_{{\rm T}_1}$), and makes the STG slightly less negative.

The remaining four molecules (B, D, E, and F) feature a 1,6-hexatriene connectivity. Due to the extended conjugation length, their S$_1$ transition energies range from 1.2–1.3 eV, which is slightly higher than in 1,4-connected systems. While B and E are permutationally related with 1,6-trans-hexatriene connectivity, D and F exhibit 1,6-cis-hexatriene connectivity. Unlike in 1,4-systems, the cis/trans effect is not very pronounced in 1,6-connected molecules. However, a trend may emerge as the cis component increases within the 1,6-connectivity.

Interestingly, both I and J, with 1,5-cis and 1,5-trans pentadiene connectivity, exhibit STG $\geq$ 0. Since the B–N separation ($R_{\rm BN}$) in 1,5-connected systems lies between that of 1,4- and 1,6-systems, this suggests that STG is not solely dictated by the B–N distance, but also influenced by through-bond resonance effects.

\paragraph{BN-helicene molecules:}

BN-helicene molecules represent a class of non-planar systems with negative STG, making them promising candidates for DFIST applications. There is growing interest in utilizing intramolecular through-space charge-transfer (CT) states in OLED design, particularly to enhance TADF efficiency\cite{schafer2024lowering,yang2024efficient,miranda2021controlling,woon2019intramolecular,rao2025recent,landi2022optimising}. However, despite this interest, photophysical studies on the excited-state dynamics and molecular conformations of such systems remain limited, underscoring the need for further investigation.

In helicenes where B and N atoms are positioned at opposite ends, the nature of the frontier MOs shifts from fully delocalized to charge-transfer-type (CT-type). In the latter case, the HOMO and LUMO densities become spatially localized in different regions of the molecule, a defining feature of S$_1$ and T$_1$ states with intramolecular CT character. 
The donor end of the molecule, where HOMO is localized lies around the B atom, while the acceptor end associated with LUMO is localized is around the N atom, implying an effective CT from B-to-N.
A similar analysis had been done for the derivatives of calicene, where through-bond CT was shown between donor and acceptor parts of the molecules without alternating HOMO/LUMO orbitals \cite{blaskovits2024singlet}.

Figure~\ref{fig:BNhelicenes} presents BNPAH molecules derived from PAHs---\#25, \#53, \#68, \#67, and \#45---exhibiting a gradual increase in helicity. We observe that as helicity increases, the STG decreases, with BN-helicene molecules based on PAHs \#67 and \#45 exhibiting negative STG. The frontier MO plots of these molecules (Figure~\ref{fig:BNhelicenes}) indicate that while excitations in PAH \#25 are fully delocalized, they gradually evolve into CT-type in helicenes. Since the localized HOMO and LUMO regions are not directly bonded through the PAH network, this suggests that a through-space CT interaction is responsible for the negative STG.

Furthermore, as shown in Figure~\ref{fig:BNhelicenes}, BN-helicenes maintain a sufficiently large B–N separation, preventing the formation of shorter B–N contacts upon relaxation and thereby preserving the CT-type nature of the frontier MOs. If B and N were too close, their interaction would lead to bond formation, suppressing CT-type behavior and preventing the emergence of negative STG.

\begin{table*}[hpt]
\centering
\caption{Excited-state properties of DFIST candidates calculated at the L-CC2/aug-cc-pVDZ level. The S$_1$ and T$_1$ excitation energies, along with singlet-triplet gaps (STG), are reported in eV. S$_1$–S$0$ oscillator strengths ($f_{0,1}$) are given in atomic units. \# refers to the index of the DFIST-BNPAH candidate, and \#PAH corresponds to the index of the parent PAH structure shown in Figure~\ref{fig:PAH}. Structures and HOMO/LUMO isosurfaces are given in Table~S5 of the SI.
}
\small 
\begin{tabular}{c c c c c c c c c c c c c r}
\hline
\#  & \# PAH & ${{\rm S}_1}$  & $f_{0,1}$  & ${{\rm T}_1}$   & STG & &
\#  & \# PAH & ${{\rm S}_1}$  & $f_{0,1}$  & ${{\rm T}_1}$   & STG \\
\hline
$1$   &  $46$  &  $1.276$  &  $0.007$  &  $1.374$  &  $-0.098$  & &  $37$  &  $32$  &  $1.081$  &  $0.010$  &  $1.101$  &  $-0.020$  \\ 
$2$   &  $21$  &  $1.383$  &  $0.002$  &  $1.468$  &  $-0.085$  & &  $38$  &  $54$  &  $1.345$  &  $0.032$  &  $1.365$  &  $-0.020$  \\ 
$3$   &  $46$  &  $1.835$  &  $0.012$  &  $1.911$  &  $-0.076$  & &  $39$  &  $31$  &  $1.978$  &  $0.003$  &  $1.997$  &  $-0.019$  \\ 
$4$   &  $46$  &  $1.748$  &  $0.013$  &  $1.821$  &  $-0.073$  & &  $40$  &  $31$  &  $2.043$  &  $0.012$  &  $2.062$  &  $-0.019$  \\ 
$5$   &  $76$  &  $1.061$  &  $0.001$  &  $1.134$  &  $-0.073$  & &  $41$  &  $49$  &  $1.222$  &  $0.013$  &  $1.240$  &  $-0.018$  \\ 
$6$   &  $6$   &  $1.914$  &  $0.012$  &  $1.983$  &  $-0.069$  & &  $42$  &  $55$  &  $1.267$  &  $0.002$  &  $1.285$  &  $-0.018$  \\ 
$7$   &  $21$  &  $1.384$  &  $0.004$  &  $1.452$  &  $-0.068$  & &  $43$  &  $57$  &  $1.832$  &  $0.028$  &  $1.850$  &  $-0.018$  \\ 
$8$   &  $21$  &  $1.685$  &  $0.011$  &  $1.753$  &  $-0.068$  & &  $44$  &  $71$  &  $1.816$  &  $0.030$  &  $1.834$  &  $-0.018$  \\ 
$9$   &  $6$   &  $1.844$  &  $0.005$  &  $1.912$  &  $-0.068$  & &  $45$  &  $49$  &  $2.188$  &  $0.005$  &  $2.205$  &  $-0.017$  \\ 
$10$  &  $21$  &  $1.601$  &  $0.004$  &  $1.668$  &  $-0.067$  & &  $46$  &  $52$  &  $1.155$  &  $0.005$  &  $1.172$  &  $-0.017$  \\ 
$11$  &  $46$  &  $1.306$  &  $0.009$  &  $1.373$  &  $-0.067$  & &  $47$  &  $37$  &  $1.158$  &  $0.003$  &  $1.170$  &  $-0.012$  \\ 
$12$  &  $46$  &  $1.683$  &  $0.005$  &  $1.749$  &  $-0.066$  & &  $48$  &  $15$  &  $1.040$  &  $0.005$  &  $1.051$  &  $-0.011$  \\ 
$13$  &  $67$  &  $1.178$  &  $0.001$  &  $1.240$  &  $-0.062$  & &  $49$  &  $64$  &  $1.102$  &  $0.001$  &  $1.113$  &  $-0.011$  \\ 
$14$  &  $46$  &  $1.810$  &  $0.013$  &  $1.869$  &  $-0.059$  & &  $50$  &  $67$  &  $1.273$  &  $0.000$  &  $1.284$  &  $-0.011$  \\ 
$15$  &  $21$  &  $1.301$  &  $0.002$  &  $1.357$  &  $-0.056$  & &  $51$  &  $26$  &  $1.381$  &  $0.017$  &  $1.391$  &  $-0.010$  \\ 
$16$  &  $21$  &  $1.919$  &  $0.008$  &  $1.973$  &  $-0.054$  & &  $52$  &  $17$  &  $1.363$  &  $0.017$  &  $1.372$  &  $-0.009$  \\ 
$17$  &  $76$  &  $1.084$  &  $0.002$  &  $1.138$  &  $-0.054$  & &  $53$  &  $26$  &  $1.367$  &  $0.020$  &  $1.376$  &  $-0.009$  \\ 
$18$  &  $21$  &  $1.890$  &  $0.010$  &  $1.942$  &  $-0.052$  & &  $54$  &  $32$  &  $1.002$  &  $0.002$  &  $1.010$  &  $-0.008$  \\ 
$19$  &  $76$  &  $1.317$  &  $0.000$  &  $1.366$  &  $-0.049$  & &  $55$  &  $36$  &  $1.368$  &  $0.037$  &  $1.376$  &  $-0.008$  \\ 
$20$  &  $21$  &  $1.270$  &  $0.000$  &  $1.316$  &  $-0.046$  & &  $56$  &  $37$  &  $1.378$  &  $0.047$  &  $1.386$  &  $-0.008$  \\ 
$21$  &  $76$  &  $1.477$  &  $0.004$  &  $1.522$  &  $-0.045$  & &  $57$  &  $69$  &  $1.433$  &  $0.048$  &  $1.441$  &  $-0.008$  \\ 
$22$  &  $76$  &  $1.370$  &  $0.006$  &  $1.414$  &  $-0.044$  & &  $58$  &  $63$  &  $1.147$  &  $0.047$  &  $1.154$  &  $-0.007$  \\ 
$23$  &  $76$  &  $1.563$  &  $0.013$  &  $1.607$  &  $-0.044$  & &  $59$  &  $61$  &  $1.455$  &  $0.027$  &  $1.461$  &  $-0.006$  \\ 
$24$  &  $46$  &  $1.330$  &  $0.000$  &  $1.373$  &  $-0.043$  & &  $60$  &  $40$  &  $1.505$  &  $0.035$  &  $1.506$  &  $-0.001$  \\ 
$25$  &  $45$  &  $1.166$  &  $0.003$  &  $1.205$  &  $-0.039$  & &  $61$  &  $49$  &  $1.279$  &  $0.010$  &  $1.280$  &  $-0.001$  \\ 
$26$  &  $46$  &  $1.028$  &  $0.006$  &  $1.064$  &  $-0.036$  & &  $62$  &  $19$  &  $1.815$  &  $0.011$  &  $1.815$  &  $0.000$   \\ 
$27$  &  $45$  &  $1.500$  &  $0.003$  &  $1.533$  &  $-0.033$  & &  $63$  &  $46$  &  $1.093$  &  $0.010$  &  $1.092$  &  $0.001$   \\ 
$28$  &  $55$  &  $1.069$  &  $0.005$  &  $1.096$  &  $-0.027$  & &  $64$  &  $54$  &  $1.462$  &  $0.052$  &  $1.460$  &  $0.002$   \\ 
$29$  &  $75$  &  $1.606$  &  $0.040$  &  $1.633$  &  $-0.027$  & &  $65$  &  $61$  &  $1.339$  &  $0.007$  &  $1.336$  &  $0.003$   \\ 
$30$  &  $46$  &  $1.324$  &  $0.000$  &  $1.349$  &  $-0.025$  & &  $66$  &  $55$  &  $1.398$  &  $0.004$  &  $1.393$  &  $0.005$   \\ 
$31$  &  $48$  &  $1.809$  &  $0.005$  &  $1.834$  &  $-0.025$  & &  $67$  &  $15$  &  $1.560$  &  $0.007$  &  $1.554$  &  $0.006$   \\ 
$32$  &  $48$  &  $1.823$  &  $0.007$  &  $1.846$  &  $-0.023$  & &  $68$  &  $71$  &  $1.776$  &  $0.007$  &  $1.768$  &  $0.008$   \\ 
$33$  &  $55$  &  $1.054$  &  $0.011$  &  $1.077$  &  $-0.023$  & &  $69$  &  $49$  &  $1.478$  &  $0.015$  &  $1.467$  &  $0.011$   \\ 
$34$  &  $38$  &  $1.216$  &  $0.011$  &  $1.238$  &  $-0.022$  & &  $70$  &  $30$  &  $1.791$  &  $0.046$  &  $1.764$  &  $0.027$   \\ 
$35$  &  $48$  &  $1.376$  &  $0.025$  &  $1.398$  &  $-0.022$  & &  $71$  &  $34$  &  $1.317$  &  $0.068$  &  $1.275$  &  $0.042$   \\ 
$36$  &  $17$  &  $1.419$  &  $0.016$  &  $1.440$  &  $-0.021$  & &  $72$  &  $34$  &  $1.319$  &  $0.075$  &  $1.242$  &  $0.077$   \\ 
\hline
\end{tabular}
\label{tab:top72}
\end{table*}

\subsection{DFIST candidates in the BNPAH chemical space: Final set after multi-Level screening\label{label:subsec3}}

For the 119 molecules that passed level-3 of the workflow with STG $< 0$ at the L-CC2/cc-pVDZ level, we carried out refined geometry optimizations and vibrational frequency analyses using the $\omega$B97X-D3/def2-TZVP method. Based on the vibrational criteria defined in Section~\ref{sec:methods}, 72 systems were confirmed to be dynamically stable, indicating that they correspond to local minima on the potential energy surface. The remaining 47 systems were found to be dynamically unstable, as their initial TPSSh/def2-SVP geometries 
eported in \RRef{chakraborty2019chemical} 
deviated significantly from the minima characterized by $\omega$B97X-D3/def2-TZVP. 
Hence, we exclude these 47 molecules from further analysis.

For the 72 stable molecules, we computed S$_1$ and T$_1$ excitation energies, singlet-triplet gaps (STG), and oscillator strengths ($f_{0,1}$) at the L-CC2/aug-cc-pVDZ level. These results are presented in Table~\ref{tab:top72}. The corresponding molecular structures and HOMO \& LUMO isosurfaces 
along with ring HOMA values are provided in Table~S5 of the SI.
These 72 molecules span 30 unique PAH scaffolds, with the most frequent contributors being:
PAH \#46 (10 molecules),
PAH \#21 (8 molecules),
PAH \#76 (6 molecules),
PAH \#49 (4 molecules each), and 
PAH \#55 (4 molecules each).

 \begin{figure*}[hpt]
    \centering
    \includegraphics[width=\linewidth]{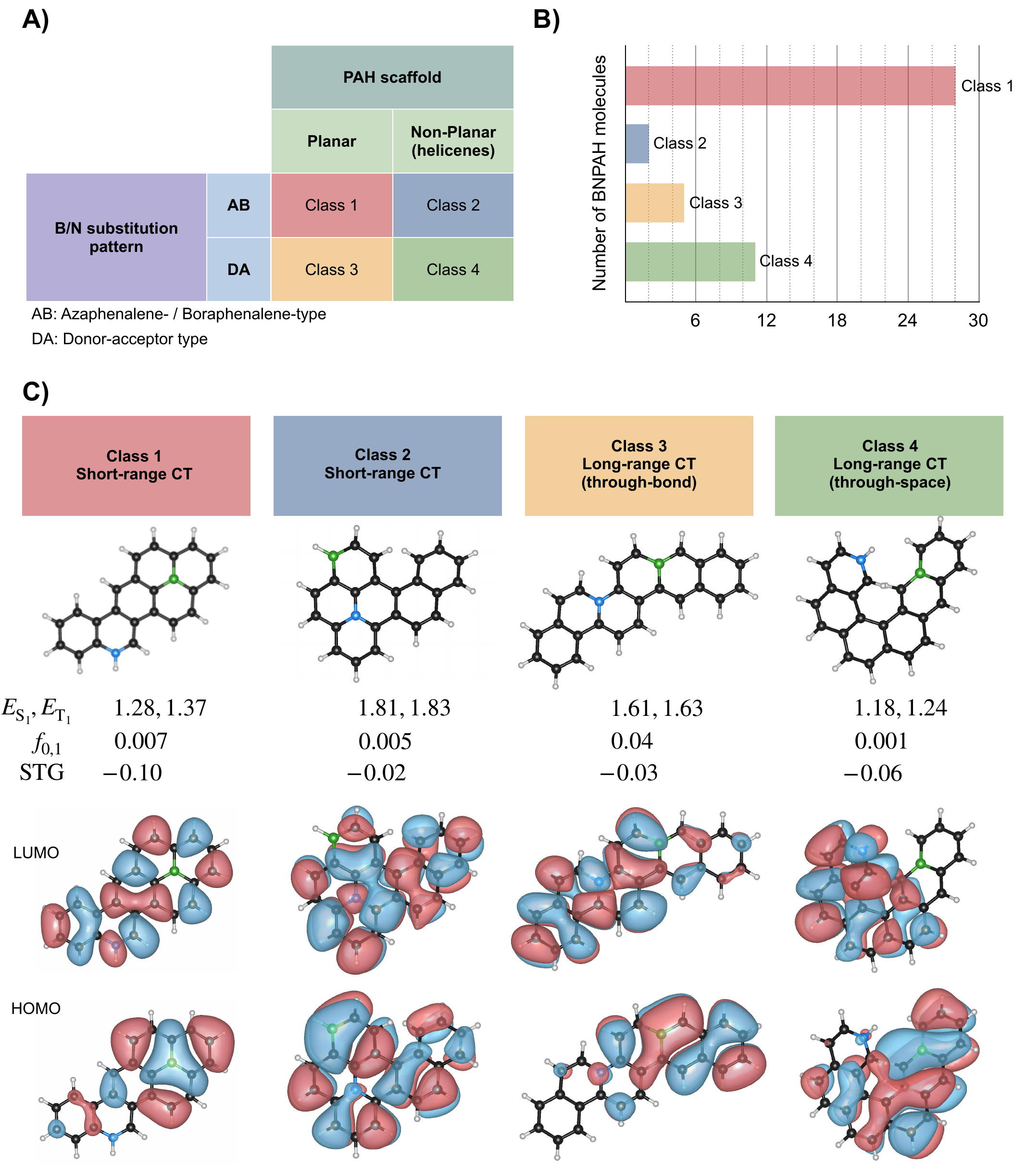}
    \caption{
    Classification of 
top 46 DFIST candidates in the BNPAH chemical space with STG$<-0.015$ eV according to L-CC2/aug-cc-pVDZ:
(A) Classification of the top candidates based on PAH scaffold characteristics and B/N substitution patterns.
(B) Bar plot showing the number of DFIST candidates in the four structural classes.
(C) The representative molecule with the most negative STG from each class is shown along with its excited-state properties. White$|$green$|$black$|$blue atoms denote H$|$B$|$C$|$N. 
\label{fig:top46}
}
\end{figure*}

Among the 72 candidates, 61 molecules exhibit negative STGs, ranging from $-$0.098 eV to $-$0.001 eV, confirming their DFIST potential. The most pronounced inversion is observed in molecule \#1 (PAH \#46) with STG = $-$0.098 eV, followed by several derivatives of the same scaffold. Only 9 molecules show STG $\geq 0$, all of which are close to zero, with the largest value being +0.077 eV.
Interestingly, several molecules with small positive STGs 
exhibit moderate oscillator strengths in the range of 0.04--0.06 a.u., suggesting more localized excitation character. However, among the 61 systems with STG $<0$, no clear correlation is observed between oscillator strength and STG. In contrast, for the 9 molecules with STG $\ge 0$, the oscillator strength tends to increase with STG, indicating a possible shift toward more delocalized or locally excited character.

As stated in \ref{label:subsec1} L-CC2/aug-cc-pVDZ has a mean error of about 0.015 eV, as established through benchmarking for 12 triangular molecules. Accordingly, STG values less than $-$0.015 eV can be considered robust, true-positive predictions within the method’s uncertainty. Based on this threshold, the first 46 of the 72 molecules in  Table~\ref{tab:top72} can be confidently classified as DFIST candidates.
The remaining 26 molecules, although exhibiting STGs within the method’s error margin, are still highly promising as TADF candidates, particularly given their small STGs. 
The shortlisted 46 DFIST BNPAH candidates with STG $<-0.015$ eV can be broadly categorized along two structural axes:
(i) the geometry of the PAH scaffold---either planar or non-planar (helicene-type), and
(ii) the B/N substitution pattern—specifically, AB-type motifs containing embedded 1AP or 1BP cores, and DA-type (donor–acceptor) motifs. 
Both substitution patterns give rise to charge-transfer (CT) character; however, AB-type systems typically exhibit short-range CT localized within a triangular subunit, while DA-type systems feature long-range CT. The nature of this CT further depends on the molecular geometry: planar PAHs support through-bond CT, whereas non-planar PAHs, such as helicenes, enable through-space CT between spatially separated B and N atoms.
This classification results in four distinct classes of DFIST candidates, as illustrated in Figure~\ref{fig:top46}.

Ring HOMA values based on $\omega$B97X-D3/def-TZVP geometries of  all 
46 BNPAH-DFIST candidates are shown in Table~S3 of the SI. In these plots, we 
find that rings containing N atoms generally have HOMA values in the range of 0.40--0.95, consistent with significant pi-electron delocalization. In contrast, B-containing rings typically show lower HOMA values, often in the range of 0.02--0.80, reflecting reduced aromatic stabilization. These trends are consistent with the observed electronic structure characteristics of the DFIST candidates, wherein localized HOMO and LUMO distributions are promoted by partial disruption of aromaticity. Thus, HOMA provides complementary structural evidence for the orbital separation central to inverted singlet–triplet gap design.

Notably, the AB-type BNPAHs, although derived from triangular PAH motifs, deviate from perfect threefold symmetry, leading to partial HOMO–LUMO overlap and thus non-vanishing oscillator strengths. While these values are lower than those seen in high-efficiency TADF emitters such as DABNA, they are nonetheless significant given the orbital separation required to achieve a negative STG. It is well known that the key electronic feature enabling inverted singlet–triplet gaps---namely, minimal spatial overlap between frontier molecular orbitals---also inherently suppresses oscillator strength \cite{aizawa2022delayed}. 
Despite this tradeoff, the BNPAH molecules reported here stand out as the only class of systems with aromatic molecular cores that exhibit negative STGs. In contrast, previously reported examples have relied on triangular molecular scaffolds ({\it e.g.}, 5AP and 7AP) or non-alternant hydrocarbon frameworks, as noted in earlier computational studies\cite{garner2023double,terence2023symmetry,garner2024enhanced,blaskovits2024excited,nigam2024artificial}.

The AB-type BNPAHs thus occupy a unique position in the landscape of DFIST emitters: their oscillator strengths are not negligible, and their emission profiles benefit from the multiresonant (MR) character\cite{madayanad2020multiresonant,kunze2024deltadft}  typical of rigid aromatic systems. These features position them as a new subclass of emitters that combine MR-like orbital separation with inverted singlet–triplet gaps. Accordingly, we refer to these as MR-DFIST emitters—molecules that unite the favorable oscillator strength and spectral sharpness of MR fluorophores with the inverted STG required for DFIST behavior.

Focusing on the top 18 DFIST candidates (those with STG $<-0.05$ eV), we find that most are derived from PAH scaffolds capable of supporting 1AP or 1BP moieties—notably, PAH \#67. However, molecule \#13, derived from a helical PAH, also appears in this group. This represents the first example of a non-planar scaffold in this study identified with negative STG, emphasizing the role of spatial orbital separation in enabling STG inversion.

\section{Conclusions\label{sec:conc}}

The DFIST landscape remains sparsely populated in terms of experimental verification. To date, 5AP and 7AP are the only systems with spectroscopically confirmed negative STGs. In this work, we present a multi-level high-throughput virtual screening framework as a complementary approach to rational molecular design, enabling the discovery of new molecular scaffolds capable of exhibiting inverted STGs. Our workflow—incorporating stringent structural stability criteria and excited-state calculations at the L-CC2 level—provides a robust platform for minimizing false-positive predictions of negative STG systems, addressing a key limitation in previous screening strategies.

By systematically exploring the BNPAH chemical space, we uncover a diverse array of DFIST candidates that extend beyond the well-known triangular frameworks of 5AP and 7AP. Starting from a dataset of 30,797 BNPAH molecules derived from exhaustive combinatorial B,\,N substitutions of 77 benzenoid PAHs, we identify 61 molecules with negative STG and 46 with STG $<-0.015$ eV 
at the L-CC2/aug-cc-pVDZ level. These low-symmetry systems exhibit non-vanishing oscillator strengths for the S$_0 \rightarrow$ S$_1$ transition, making them promising emitters. Additionally, our dataset includes several hundred BNPAH molecules with near-zero STGs, which could enable temperature-independent TADF through thermally accessible RISC.

Our analysis reveals that these 46 DFIST candidates can be categorized into four distinct structural classes, based on a combination of PAH topology (planar vs. non-planar) and heteroatom substitution pattern (1AP/1BP-like vs. donor–acceptor-like). This classification highlights how molecular geometry and substitution pattern jointly influence the STG, offering new handles for design and optimization. 
Importantly, we observe that some of the low-STG prototypes are structurally stable only in their fused PAH form, despite the instability of their isolated B,N subunits. For example, 1BP is unstable as an isolated unit but becomes stable—and retains a negative STG—when embedded in a larger PAH scaffold. This finding underscores the potential of extended conjugation as a stabilizing design principle. We also identify non-planar BN-helicenes as promising DFIST candidates. These systems feature through-space charge transfer character, with spatially separated HOMO and LUMO densities facilitating STG inversion. However, their non-planar geometry leads to limited stacking interactions, which may reduce exciton diffusion and energy transfer efficiency in solid-state applications.

Although we limit our exploration to BNPAHs containing only one B and one N atom, the complete BNPAH chemical space---based on 77 PAHs---with same number of B and N atoms
includes over 7.4 trillion unique molecules (7.4×10\textsuperscript{12})\cite{chakraborty2019chemical}. Exploring such a vast space will require the integration of evolutionary algorithms, generative models, and other machine-learning-driven strategies.
Finally, while ring-closing synthetic techniques can be used to incorporate B,\,N units into aromatic frameworks\cite{wang2015bn,chen2022multiple,gotoh2021syntheses,appiarius20211,jaye2017implications}, synthetic access to specific substitution patterns remains challenging\cite{xu20111,giustra2018state}. 
The 46 DFIST candidates identified here provide concrete molecular targets for synthesis and should be subjected to further refinement using high-level theoretical methods and experimental validation.

\section{Supplementary Information}
i) Names and SMILES representations of the 77 parent PAHs used to construct the BNPAH chemical space;
ii) Theoretical best estimates, along with CC2 and L-CC2 excited-state energies (S$_1$, T$_1$, and STG) for twelve triangular benchmark systems;
iii) Structures, frontier molecular orbitals (HOMO and LUMO), and ring aromaticity (HOMA indices) for the 72 shortlisted BNPAH-DFIST candidates.

\section{Data Availability}
The data that support the findings of this study are
within the article and its supplementary material.

\section{Acknowledgments}
We acknowledge the support of the 
Department of Atomic Energy, Government
of India, under Project Identification No.~RTI~4007. 
All calculations have been performed using 
the Helios computer cluster, 
which is an integral part of the MolDis 
Big Data facility, 
TIFR Hyderabad \href{http://moldis.tifrh.res.in}{(http://moldis.tifrh.res.in)}.

\section{Author Declarations}

\textbf{AM}: 
Conceptualization (lead); 
Analysis (lead); 
Data collection (lead); 
Writing -- original draft (lead); 
Writing -- review and editing (equal).
\textbf{SD}: 
Analysis (supporting); 
Data collection (supporting); 
Writing -- original draft (supporting); 
Writing -- review and editing (equal).
\textbf{RR}: 
Conceptualization (lead); 
Analysis (lead); 
Data collection (supporting); 
Funding acquisition; 
Project administration; 
Supervision; Resources; 
Writing -- original draft (lead); 
Writing -- review and editing (equal).

\subsection{Conflicts of Interest}
The authors have no conflicts of interest to disclose.

\section{References}
\bibliography{ref} 
\end{document}